\documentclass{mem}
\usepackage{natbib}\usepackage{txfonts}\usepackage{balance}
\usepackage{graphicx}
\usepackage[a4paper,breaklinks,dvipdfm]{hyperref}
\idline{000}{001}
\begin{document}
\def\teff{$T\rm_{eff }$}
\def\kms{$\mathrm {km s}^{-1}$}

\title{
Probing Unification Scenarios with Atomic Clocks
}

   \subtitle{}

\author{
M.D. \,Juli\~ao\inst{1,2}, M. C. \,Ferreira\inst{1,3}, C. J. A. P. \,Martins\inst{1} \and A. M. R. V. L. \,Monteiro\inst{1,3,4}
          }

  \offprints{M. Juli\~ao}

\institute{
Centro de Astrof\'{\i}sica, Universidade do Porto, Rua das Estrelas, 4150-762 Porto, Portugal
\and
Faculdade de Engenharia, Universidade do Porto, Rua Dr Roberto Frias, 4200-465 Porto, Portugal
\and
Faculdade de Ci\^encias, Universidade do Porto, Rua do Campo Alegre, 4150-007 Porto, Portugal
\and
Department of Applied Physics, Delft University of Technology, P.O. Box 5046, 2600 GA Delft, The Netherlands
}

\authorrunning{Juli\~ao et al.}

\titlerunning{Probing Unification with Atomic Clocks}

\abstract{ We make use of the comparison between measurements of various pairs of atomic clocks to impose constraints on coupled variations of fundamental constants in severs  unification scenarios. We obtain null results for the proton-to-electron mass ratio ${\dot\mu}/{\mu}=(0.68\pm5.79)\times10^{-16}\, {\rm yr}{}^{-1}$ and for the gyromagnetic factor ${\dot g_p}/{g_p}=(-0.72\pm0.89)\times10^{-16}\, {\rm yr}{}^{-1}$ (both of these being at the $95\%$ confidence level). These results are compatible with theoretical expectations on unification scenarios (which we briefly describe), but much freedom exists due to the presence of a degeneracy direction in the relevant parameter space.

}
\maketitle{}

\section{Introduction}

Scalar fields help us to understand several crucial aspects in fundamental physics and cosmology, such as the LHC evidence for the Higgs particle, inflation, cosmological phase transitions or dynamical dark energy. As there is, as yet, no evidence for their presence in cosmological contexts, looking for them (or setting constraints on scenarios invoking them) is an extremely exciting and topical quest.

Probably the most direct way to look for them is to search for spacetime variations of nature's fundamental constants \citep{cjm202,gb207,uz211}. It is well known that fundamental couplings run with energy, and there are many classes of models in particle physics and cosmology in which fundamental couplings are spacetime-dependent. The fine structure constant $\alpha$ and the proton-to-electron mass ratio $\mu$ are ideal probes, and can currently be measured up to redshifts $ z \sim4$.

As was amply discussed in this meeting (and is reported elsewhere in these proceedings), the observational status of these searchers is intriguing. There are astrophysical measurements suggesting variations \citep{mu204,rh206,web211} while others with comparable sensitivity find null results \citep{sr97,ki208,pm97}. One of the goals of the UVES Large Program for Testing Fundamental Physics \citep{LP1,LP2} is precisely to clarify this issue.

Meanwhile, there are also several laboratory and astrophysical measurements that are sensitive to various products of dimensionless couplings, rather than to each of them separately. Moreover, by combining several of these measurements it is often possible to break some of the associated degeneracies, as well as to derive constraints on particular classes of models of fundamental physics and unification \citep{FJMM1,FJMM2}.

In what follows we briefly describe the results of the application of this methodology to current atomic clock measurements. A.M. Monteiro's contribution to these proceedings describes an analogous study for the case of a particular astrophysical system.

\section{Constraints from Atomic Clocks}

Mechanical clocks work by counting the vibrations of something which has a sufficiently stable frequency (at the scale it is designed to work), $\nu$. For instance, in a pendulum clock, the frequency is a function of several parameters: 

\begin{equation}
\nu=\nu(temperature, pressure, gravity, ...) .
\end{equation}

In atomic clocks, the characteristic frequency corresponds to the frequency of some transition. In particular, for an atomic clock of an alkali-like element, that frequency is the hyperfine frequency of the element---which depends, again, on several quantities (\citep{luo211}

\begin{equation}
\nu_{hfs}=\nu_{hfs}(R_{\infty}c,A_{hfs},g_{i},\alpha^{2},\mu, F_{hfs}(\alpha)) ,
\end{equation}

\noindent where $R_{inf}$ is the Rydberg constant, $A_{hfs}$ is a numerical factor depending on the atomic species,
$g_{i}=2\mu_{i}/\mu_{N}$ is the gyromagnetic factor ($\mu_{i}$ being nuclear magnetic moment and $\mu_{N}=e/2m_{p}$ the nuclear magneton), $\alpha$ is the fine structure constant, $ \mu \equiv \frac{m_{p}}{m_{e}} $ is the proton-to-electron mass ratio and $F_{hfs}(\alpha)$ accounts for relativistic corrections.

By comparing clocks whose frequencies have different sensitivities to parameters such as $\alpha$, $\mu$  and $g_i$, we can therefore obtain constraints on the drift rate of the relevant parameter combinations. Schematically we'll typically have a relation of the form

\begin{equation}
\frac{\delta \nu_{AB}}{\nu_{AB}}= \lambda_{ g_{p}}\frac{\delta g_{p}}{g_{p}} + \lambda_{ g_{n}}\frac{\delta g_{n}}{g_{n}} + \lambda_{b}\frac{\delta b}{b} + \lambda_{\mu}\frac{\delta \mu}{\mu} + \lambda_{\alpha}\frac{\delta \alpha}{\alpha}
\end{equation}

By combining several measurements with different sensitivity coefficients, we can obtain constraints on the drifts of the individual quantities. And example of this type of analysis is \citet{luo211} and we have recently updated this using newer data \citep{FJMM1}. In what follows we summarise these results.

\begin{figure}
\includegraphics[width=0.5\textwidth]{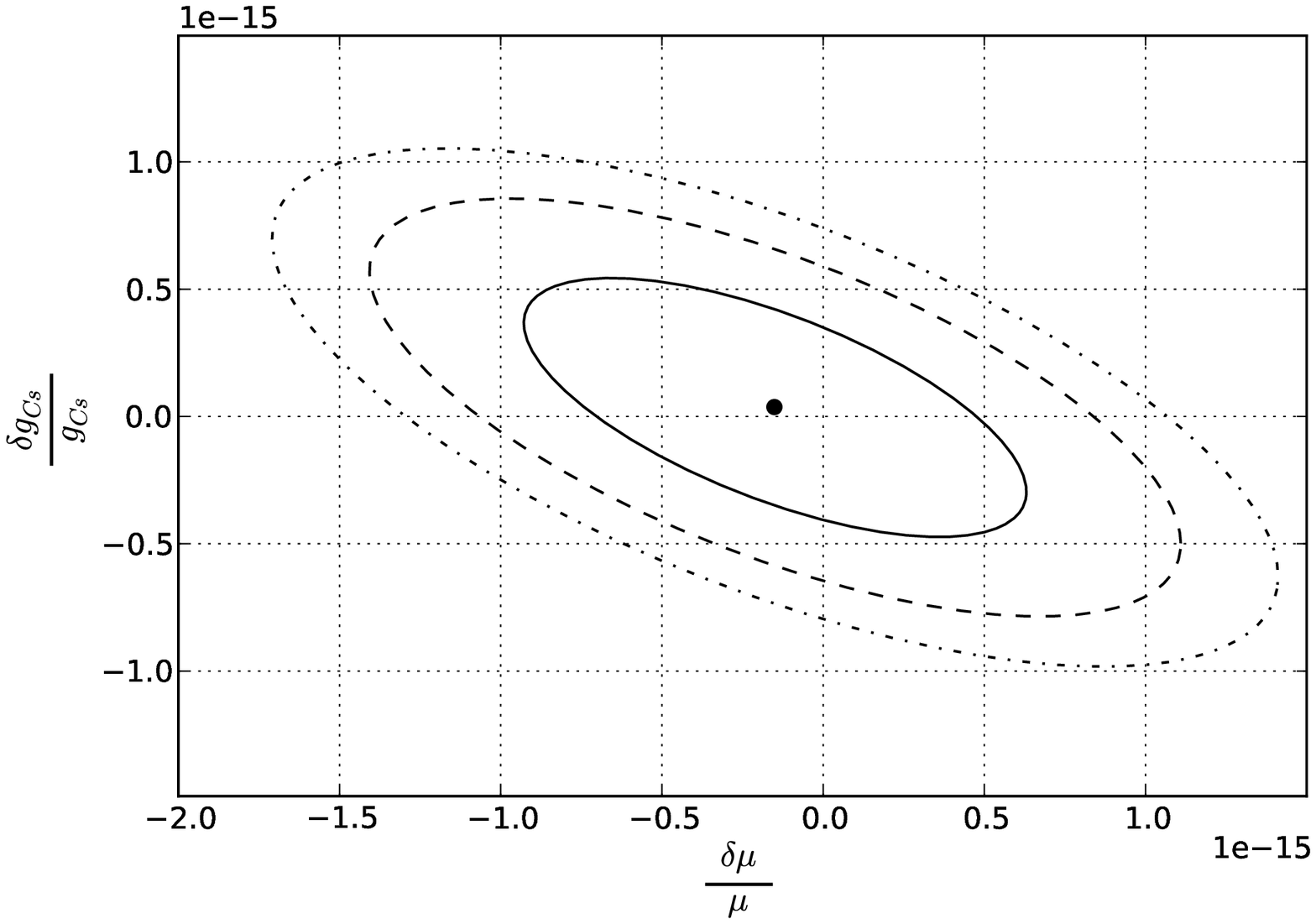} 
\includegraphics[width=0.5\textwidth]{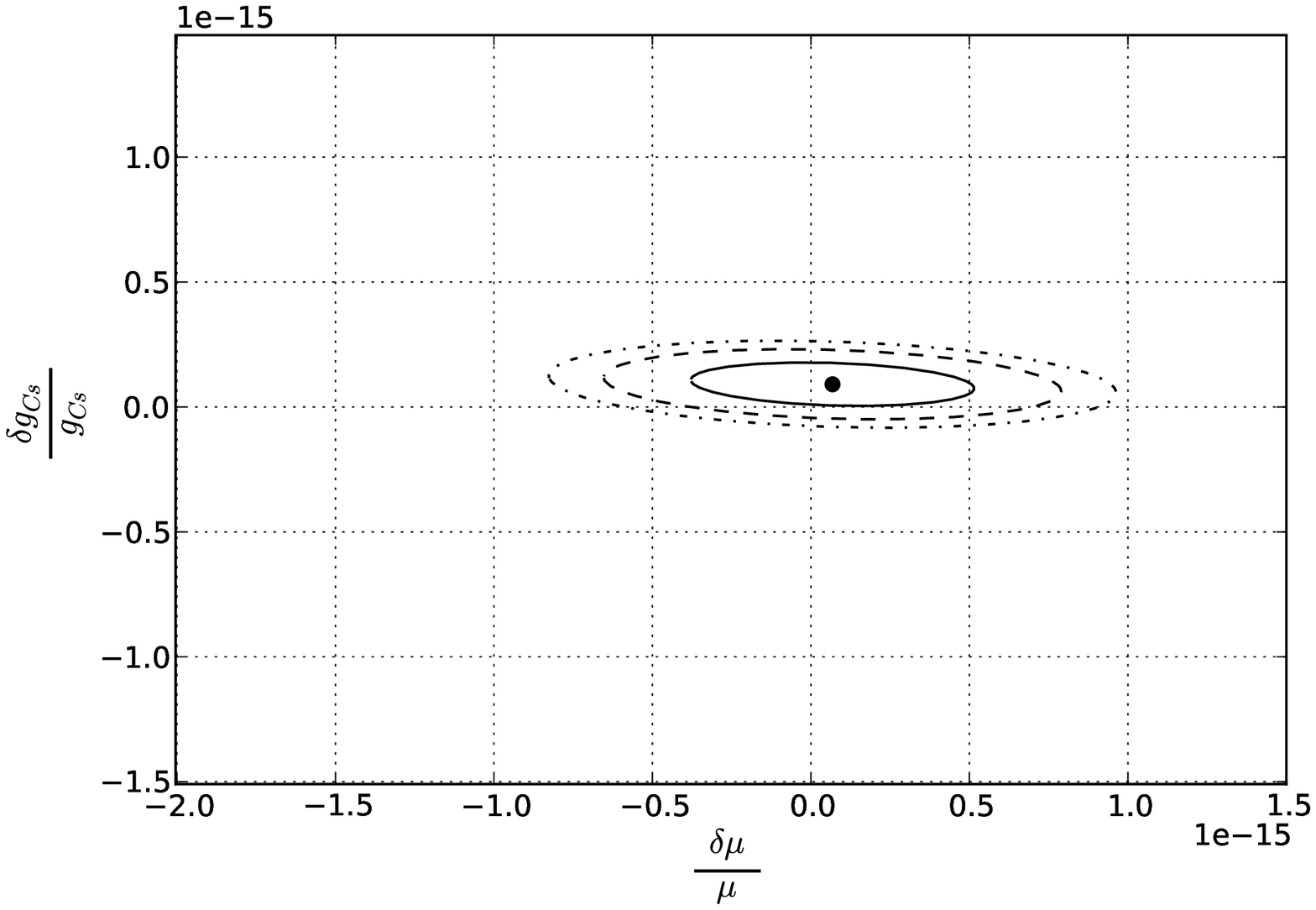} 
\caption{\footnotesize
Atomic clock constraints on the $\mu$-$g_{Cs}$ parameter space. The top panel shows the constraints obtained with the data discussed in \protect\citet{luo211}, while the bottom panel shows the constraints derived from the most recent data, i.e. using \protect\citet{pk210}, \protect\citet{gu212} instead of \protect\citet{pk204}, \protect\citet{bz205}. In both cases the one-, two- and three-sigma likelihood contours are plotted. Notice the change in the degeneracy direction. Reprinted, with permission, from \citet{FJMM1}.}
\label{mu-gcs}
\end{figure}

The experimental data is listed in Table \ref{table1}, and the resulting one-, two-, and three-sigma likelihood contours in the $\mu$-$g_{Cs}$ plane are plotted in Fig. \ref{mu-gcs}. Note that since thee first clock comparison (Hg-Al) gives a direct and very tight constraint on $\alpha$, it is legitimate to use this data set to impose bounds in the $\mu - g_{cs}$ space.

\begin{table*}
\begin{center}
\begin{tabular}{cccc}
\hline
 Clocks & $\nu_{AB}$ & ${\dot \nu_{AB}}/{\nu_{AB}}$ (yr${}^{-1}$) & Ref. \\ 
 \hline
 \hline
Hg-Al & $\alpha^{-3.208}$ & $(5.3\pm7.9)\times10^{-17}$ & \citet{rs208}  \\
\hline
Cs-SF${}_6$ & $g_{Cs}\mu^{1/2}\alpha^{2.83}$ & $(-1.9\pm0.12_{sta}\pm2.7_{sys})\times10^{-14}$ & \citet{sh208} \\
Cs-H & $g_{Cs}\mu\alpha^{2.83}$ & $(3.2\pm6.3)\times10^{-15}$ & \citet{fs204} \\
Cs-Sr & $g_{Cs}\mu\alpha^{2.77}$ & $(1.0\pm1.8)\times10^{-15}$ & \citet{bl208}  \\
Cs-Hg & $g_{Cs}\mu\alpha^{6.03}$ & $(-3.7\pm3.9)\times10^{-16}$ & \citet{ft207} \\
\hline
Cs-Yb & $g_{Cs}\mu\alpha^{1.93}$ & $(0.78\pm1.40)\times10^{-15}$ & \citet{pk204} \\
Cs-Rb & ($g_{Cs}/g_{Rb})\alpha^{0.49}$ & $(0.5\pm5.3)\times10^{-16}$ & \citet{bz205} \\
\hline
Cs-Yb & $g_{Cs}\mu\alpha^{1.93}$ & $(0.49\pm0.41)\times10^{-15}$ & \citet{pk210} \\
Cs-Rb & ($g_{Cs}/g_{Rb})\alpha^{0.49}$ & $(1.39\pm0.91)\times10^{-16}$ & \citet{gu212} \\
\hline
\end{tabular}
\caption{\label{table1} Atomic clock constraints of varying fundamental couplings. The second column shows the combination of couplings to which the clock comparison is sensitive, and the third column shows the corresponding experimental bound. The measurements in the first seven lines were the ones used in \citet{luo211}; in our analysis the limits from Rubidium and Ytterbium clocks (lines 6 and 7) have been updated to those in lines 8 and 9.}
\end{center}
\end{table*}

From this combined analysis we calculated the 95\% confidence intervals for both parameters

\begin{equation}
\frac{\dot \mu}{\mu}=(6.8  \pm 57.6) \times 10^{-17} yr^{-1}
\end{equation}

\begin{equation}
\frac{\dot g_{Cs}}{g_{Cs}}=(9.1 \pm 11.3) \times 10^{-17} yr^{-1}
\end{equation}

\noindent or equivalently

\begin{equation}
\frac{\dot g_{p}}{g_{p}}=(-7.2 \pm 8.9) \times 10^{-17} yr^{-1}
\end{equation}

These should be compared to the value obtained by \citet{rs208} :

\begin{equation}
\frac{\dot \alpha}{\alpha}=(-1.7 \pm 4.9) \times 10^{-17} yr^{-1}
\end{equation}

Our results are therefore consistent with the stability of the constants at the probed level of precision. We note that there is a more recent measurement for the variation of $\alpha$ \citep{lee213}, but it has very little effect on our results, since the bound of \citet{rs208} is much tighter.

\section{Phenomenology of Unification}

We can go beyond the previous analysis, by using the data to constrain a class of unification models with simultaneous but related variations of several fundamental couplings---specifically, for our present purposes, $\alpha$, $\mu$ and $g_{p}$ - that are model dependent.

We will follow the approach of \citet{coc207} and \citet{luo211}, considering a class of grand unification models in which the weak scale is determined by dimensional transmutation, and assuming that relative variation of all the Yukawa couplings is the same. As in \citet{co95}, we assume that the variation of the couplings is driven by a dilation-type scalar field. With these assumptions one can show \citep{coc207} that $\mu$ and $\alpha$ are related by

\begin{equation}
\frac{\Delta \mu}{\mu}=[0.8R - 0.3(1+S)]\frac{\Delta \alpha}{\alpha},
\end{equation}

where $R$ and $S$ can be taken as free phenomenological parameters. On the other hand, for the proton and neutron $g$-factors one has \citep{flsn,fl204,fl206}

\begin{equation}
\frac{\Delta g_p}{g_p}=[0.10R-0.04(1+S)]\frac{\Delta\alpha}{\alpha}\,
\end{equation}

\begin{equation}
\frac{\Delta g_n}{g_n}=[0.12R-0.05(1+S)]\frac{\Delta\alpha}{\alpha}\,,
\end{equation}

\noindent which together allow us to map any measurement of a combination of constants into a constraint on the (R,S,$\alpha$) parameter space.

Given the theoretical uncertainty, there are two possible approaches to relate the various parameters, further discussed in \citet{luo211}. In a simple shell model:

\begin{equation}\label{grb1}
\frac{\Delta g_{Rb}}{g_{Rb}} \simeq0.736 \frac{\Delta g_p}{g_p}\simeq[0.07R-0.03(1+S)]\frac{\Delta\alpha}{\alpha}\,
\end{equation}

\begin{equation}\label{gcs1}
\frac{\Delta g_{Cs}}{g_{Cs}}\simeq-1.266\frac{\Delta g_p}{g_p}\simeq[-0.13R+0.05(1+S)]\frac{\Delta\alpha}{\alpha}\,
\end{equation}

\noindent while with a more accurate phenomenological description, motivated from experimental results and including a dependence on $g_n$ and the spin-spin interaction, 

\begin{equation}\label{grb2}
\frac{\Delta g_{Rb}}{g_{Rb}}\simeq[0.014R-0.007(1+S)]\frac{\Delta\alpha}{\alpha}\,
\end{equation}

\begin{equation}\label{gcs2}
\frac{\Delta g_{Cs}}{g_{Cs}}\simeq[-0.007R+0.004(1+S)]\frac{\Delta\alpha}{\alpha}\,. 
\end{equation}

\begin{figure}
\includegraphics[width=0.5\textwidth]{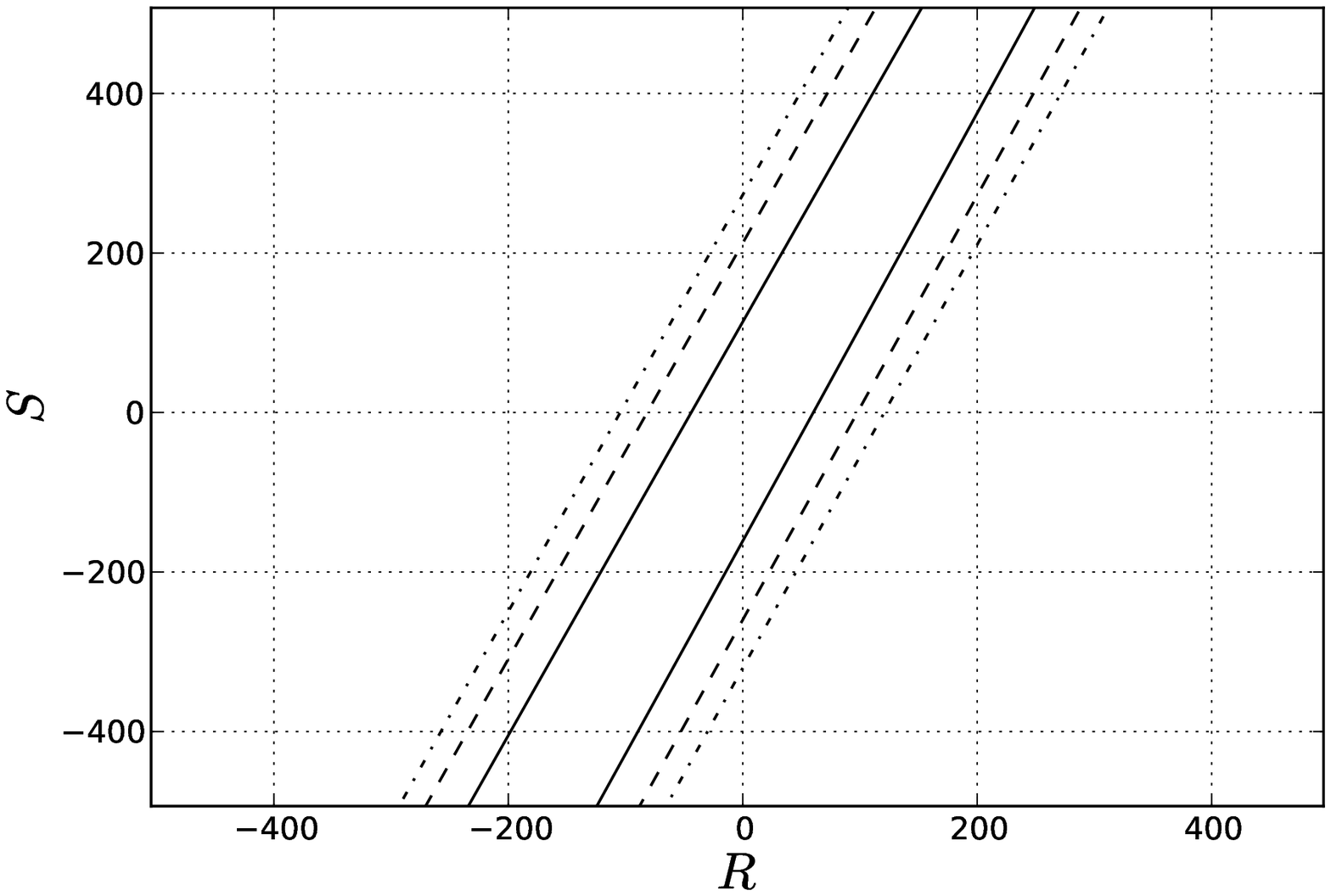} 
\includegraphics[width=0.5\textwidth]{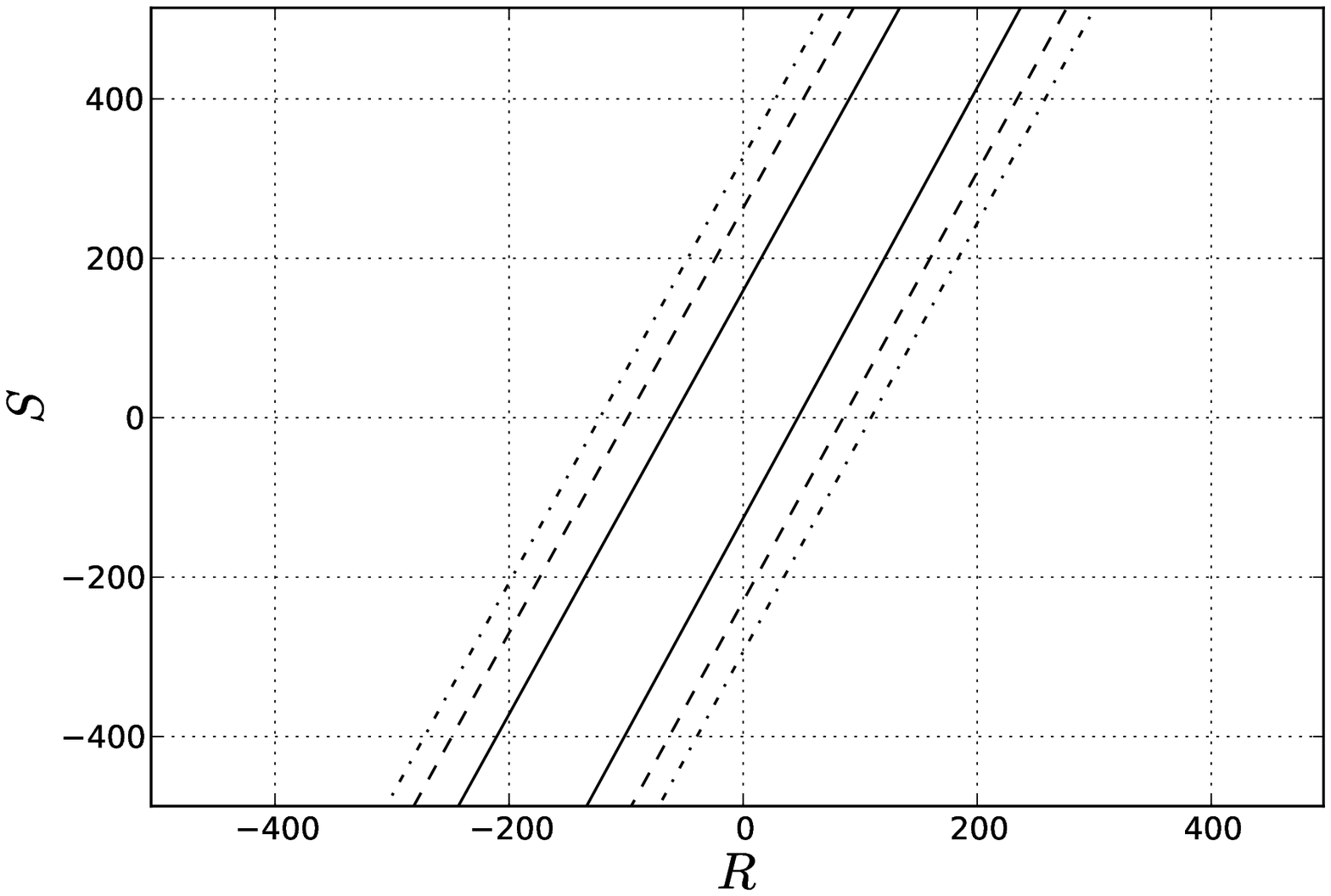} 
\caption{\footnotesize
Atomic clock constraints on the $R-S$ parameter space, using the same data as for the bottom panel of \protect\ref{mu-gcs}. The one-, two- and three-sigma likelihood contours are plotted. In the top panel the relations between the variations of the gyromagnetic rations and $\alpha$ are the ones in the simple shell model, while the bottom panel assumes the relations in the more accurate phenomenological description. Reprinted, with permission, from \citet{FJMM1}.}
\label{RandS}
\end{figure}

The results of this analysis, with both sets of assumptions, are shown in Fig. \ref{RandS}; one can see that given the current experimental uncertainties the theoretical ones are not a limiting factor (however, they may become a limiting factor as the experimental results improve). The degeneracy direction can be characterised by

\begin{equation}
(S+1) - 2.7R = -5 \pm 15
\end{equation}

Given simple theoretical expectations on unification scenarios, one may naively expect typical values for $R$ and $S$, say $R \sim 30$ and $S \sim 160$. So, fixing these two separately, one can find the following bounds for the other: 

\begin{equation}
R=61\pm 71,\qquad {\rm for} $S=160$
\end{equation}

\begin{equation}
S=76\pm 197,\qquad {\rm for} $R=30$;
\end{equation}

\noindent we note that In both methods of calculation these values are in agreement, at the 95 \% confidence level. As one may expect from the fact that the experimental values are consistent with no variations, these  results are compatible with theoretical expectations on unification scenarios.

\section{Conclusions}

We have considered the latest tests of the stability of nature’s fundamental constants using atomic clocks and
discussed their usage as a tool to constrain unification scenarios. A global analysis of existing measurements,
assuming the tight bound of \citet{rs208}, allows us to obtain separate updated constraints on $\mu$ and $g_p$.

We then used these same measurements to set constraints on a simple but generic phenomenological description of unification scenarios. The atomic clock results lead to a degeneracy in the space of the two relevant phenomenological parameters ($R$ and $S$). This degeneracy may be broken by measurements in astrophysical systems that have different sensitivities to these parameters. An example is provided in A.M. Monteiro's contribution to these proceedings.

\begin{acknowledgements}
We acknowledge the financial support of grant PTDC/FIS/111725/2009 from FCT (Portugal). CJM is also supported by an FCT Research Professorship, contract reference IF/00064/2012. We would also like to sincerely acknowledge the staff of the Sesto Center for Astrophysics for the way we were welcomed.
\end{acknowledgements}

\bibliographystyle{aa}

\end{document}